\documentclass[aps,twocolumn,10pt,secnumroman,amsmath,amssymb,nofootinbib]{revtex4}

\pdfoutput=1

\usepackage{graphics}      % standard graphics specifications
\usepackage{graphicx}      % alternative graphics specifications
\usepackage{longtable}     % helps with long table options
\usepackage{url}           % for on-line citations
\usepackage{bm}            % special 'bold-math' package
\usepackage{array}   
\usepackage{float}
\usepackage{setspace}
\usepackage{algorithm}
\usepackage{algpseudocode}
\usepackage{listings}

\usepackage{fancyhdr}
\begin{document}
\title {Numerical and analytical approaches to an advection-diffusion problem at small Reynolds number and large P\'eclet number}
\author {Nathaniel J. Fuller}
\email[email address: ]{nfuller@umich.edu}
\affiliation {Department of Natural Sciences, University of Michigan-Dearborn, Dearborn, Michigan 48128}
\author {Nicholas A. Licata}
\email[email address: ]{licata@umich.edu}
\affiliation {Department of Natural Sciences, University of Michigan-Dearborn, Dearborn, Michigan 48128}
\date{\today}

\begin{abstract}
Obtaining a detailed understanding of the physical interactions between a cell and its environment often requires information about the flow of fluid surrounding the cell.  Cells must be able to effectively absorb and discard material in order to survive.  Strategies for nutrient acquisition and toxin disposal, which have been evolutionarily selected for their efficacy, should reflect knowledge of the physics underlying this mass transport problem.  Motivated by these considerations, in this paper we discuss the results from an undergraduate research project on the advection-diffusion equation at small Reynolds number and large P\'eclet number.  In particular, we consider the problem of mass transport for a Stokesian spherical swimmer.  We approach the problem numerically and analytically through a rescaling of the concentration boundary layer.  A biophysically motivated first-passage problem for the absorption of material by the swimming cell demonstrates quantitative agreement between the numerical and analytical approaches.  We conclude by discussing the connections between our results and the design of {\it smart} toxin disposal systems.  
\end{abstract}

\maketitle

\section{Introduction}
The survival of any microorganism is dependent on its ability to control the movement of nutrients and waste between the cell interior and the external environment. Nutrients must be easily captured and waste easily disposed with minimal metabolic exertion \cite{cole2013cost}. Factors which determine the efficiency of any material transfer mechanism for gathering and disposing of small molecules include the microscale flow profile surrounding the cell \cite{short2006flows}, the diffusion coefficient of molecules being transported, and the location of transport receptors on the cell surface \cite{whalen2012actin,Song}.  As with any other trait, the physical architecture of the transport receptor arrangement is evolutionarily selected to work efficiently within the particular environment that the organism lives. A more detailed understanding of the mass transport problem for molecules in the extracellular fluid can yield valuable information for discerning why a given architecture is preferable over other possible choices \cite{vogel1994life}.  

An example which motivates the current study is understanding the mechanism of toxin disposal by sea urchin embryos \cite{whalen2012actin,gokirmak2012localization}.  Early in development, microvilli (microscopic cylindrical cell membrane protrusions) on the embryo grow to several microns in length, with transport receptors localized at the tips of the microvilli.  This localization may help facilitate more efficient toxin disposal by the embryo.  By releasing molecules away from the body of the cell, the chances are reduced that expelled toxins are reabsorbed.  In this setting, the mass transport problem for molecules in the extracellular fluid is described an advection-diffusion equation at small Reynolds number and large P\'eclet number.  Interestingly, the characteristic lengthscale for the concentration boundary layer may provide a physical rationale for the length of the microvilli involved in the toxin transport  \cite{urchin}.  

Calculating the efficiency of a material transfer mechanism requires the solution of two distinct problems. The first problem is to determine the flow profile of the extracellular fluid.  In the present context, the Reynolds number of the flow is much smaller than one, which justifies the use of the Stokes equation for describing microscale flows around individual cells \cite{felderhof2015stokesian}.  The second problem is to find a solution of the advection-diffusion equation using the calculated flow profile as input. The solution of the advection-diffusion equation provides the concentration profile governing the average motion of particles dissolved in the fluid. Knowledge of the concentration profile leads directly to information about the probability for a given particle to contact the outer envelope of a cell. Since the absorption of a particle requires that it first make contact with the cell surface, the concentration profile can be used to find a theoretical maximum absorption probability \cite{berg1977physics}.

Although the equations which govern the extracellular flow and diffusion of small molecules are well understood, deriving their solutions or even their approximate solutions is often quite involved \cite{ifm,ifac,acrivos1960solution, acrivos1962heat, acrivos1965asymptotic,goddard1966asymptotic}. The difficulties faced in finding accurate analytic solutions to the advection-diffusion equation can make numerical schemes an attractive method for finding concentration profiles \cite{cfmht,ecfd,icfme}.  In this paper we present both numerical solutions, and an analytic approximation based on a rescaling of the concentration boundary layer.  The discussion of the analytic approximation is self-contained and does not assume any familiarity with the associated boundary layer methods in fluid mechanics.  The pedagogical exposition is intended so that an advanced undergraduate student with exposure to boundary value problems common in a quantum mechanics or electricity and magnetism course can follow the derivation.  

In what follows we consider the flow profile around a single sphere due to deformations of its surface, which is a common model for active particles \cite{felderhof2015stokesian}.  This model is chosen for several reasons. First, it provides a model to mimic the flow around a spherical cell swimming through a liquid. Second, since the solution for the flow profile is known, we can make progress on an analytic approximation to the associated advection-diffusion equation.  Our analysis of the concentration profile focuses on the first-passage probability of a particle in the extracellular fluid being absorbed by the cell \cite{redner2001guide}. The first-passage probability reveals the theoretical maximum efficiency for the capture of small molecules by the cell \cite{berg1977physics}.

\section{Fluid Flow}
Consider a spherical cell of radius $R$ in a fluid.  We assume the fluid is infinite in extent and also at rest at infinity. In the case of present interest, the Reynolds number corresponding to the flow around the cell is generally much less than one. At small Reynolds numbers, the magnitude of the inertial terms in the Navier-Stokes equation become much less than the magnitude of the viscous terms. The disparity in magnitude allows for a linearization of the Navier-Stokes equation into the Stokes equation, which neglects the inertial terms.  

The flow velocity $\bm{v}$ satisfies the Stokes equation
\begin{equation}
\mu\nabla^{2}\bm{v}-\nabla p = 0
\end{equation}
where $p$ is the pressure profile and $\mu$ is the dynamic viscosity of water. Since the speed of the flow will be much less than the speed of sound in water, the fluid is considered to be perfectly incompressible and satisfy the incompressibility condition
\begin{equation}
\nabla\cdot\bm{v} = 0.
\end{equation}
The fluid is set into motion through axially symmetric deformations of the spherical surface of the cell.  At small Reynolds number the fluid motion is then determined by the no-slip boundary condition at the surface.  The details for the calculation of the flow velocity can be found in \cite{felderhof2015stokesian}.  Defining the dimensionless length $\xi=r/R$, in the instantaneous rest frame of the cell, the dimensionless fluid velocity is $\bm{u}(\bm{\xi},t) = \bm{v}(\bm{\xi},t)/U(t)$ where 
\begin{eqnarray}
\bm{u}(\bm{\xi},t) = -\bm{\hat{z}} + \sum_{\ell=1}^{\infty} m_{\ell}(t) \bm{u}_{\ell}(\xi,\theta) + \sum_{\ell=2}^{\infty} k_{\ell}(t) \bm{v}_{\ell}(\xi,\theta)
\label{flowfield}
\end{eqnarray}
and $\bm{U}(t)=U(t)\bm{\hat{z}}$ is the translational velocity of the cell.  The defining relations for $\bm{u}_{\ell}(\xi,\theta)$ and $\bm{v}_{\ell}(\xi,\theta)$ are 
\begin{eqnarray}
\bm{u}_{\ell}(\xi,\theta) = \xi^{-(\ell+2)} \left( (\ell+1)P_{\ell}(\cos \theta) \bm{\hat\xi}  + P_{\ell}^{1}(\cos \theta) \bm{\hat{\theta}}\right) \\
\bm{v}_{\ell}(\xi,\theta) = \xi^{-\ell} \left( (\ell+1)P_{\ell}(\cos \theta) \bm{\hat\xi}  + \frac{(\ell-2)}{\ell}P_{\ell}^{1}(\cos \theta) \bm{\hat{\theta}}\right)
\end{eqnarray}
with $P_{\ell}(\cos \theta)$ the Legendre polynomial and $P_{\ell}^{1}(\cos \theta)$ the associated Legendre function.  

\section{Numerical Model}
With a solution for the flow profile in hand, we now turn to the second problem, determining the concentration profile from the advection-diffusion equation.  Our eventual aim is to find the first passage probability of a dissolved molecule released at position ($\xi',\theta',\phi'$) through the inner boundary at $\xi=1$. To accomplish this, we solve the advection-diffusion equation using the known velocity profile. In this section we first discuss a numerical solution of the problem.  Later we will compare our results to an analytic approximation.  

For an incompressible fluid, the advection-diffusion equation reads as
\begin{equation}
\frac{\partial C}{\partial t}+\bm{v}\cdot\nabla C=D\nabla^{2}C
\label{ade}
\end{equation}
where $D$ is the diffusion coefficient for the molecule of interest, $C$ is the molecular concentration, and $t$ is time. We define the dimensionless concentration $c=CR^{3}$ and dimensionless time $\tau=(Dt)/R^{2}$.  In what follows we will work with a time independent flow velocity $\overline{\bm{v}}$ by considering a temporal average over a period of the fast swimming motion, in which case the P\'eclet number can be defined as $\mathrm{Pe}=(R\overline{U})/D$.  The dimensionless form of Eq. (\ref{ade}) in spherical polar coordinates is
\begin{widetext}
\begin{eqnarray}
\frac{\partial c}{\partial\tau}+\mathrm{Pe}\left( -\cos \theta +\sum_{\ell=1}^{\infty} \overline{m}_{\ell} \xi^{-(\ell+2)} (\ell + 1)P_{\ell}(\cos \theta) + \sum_{\ell=2}^{\infty} \overline{k}_{\ell} \xi^{-\ell} (\ell + 1)P_{\ell}(\cos \theta) \right) \frac{\partial c}{\partial \xi} \nonumber \\
+ \mathrm{Pe}\left( \sin \theta +\sum_{\ell=1}^{\infty} \overline{m}_{\ell} \xi^{-(\ell+2)}P_{\ell}^{1}(\cos \theta) + \sum_{\ell=2}^{\infty} \overline{k}_{\ell} \xi^{-\ell} \frac{(\ell - 2)}{2}P_{\ell}^{1}(\cos \theta)\right)\frac{1}{\xi}\frac{\partial c}{\partial \theta} = \nabla^{2}c.
\label{dade}
\end{eqnarray}
\end{widetext}
Building the numerical model is generally simpler and less prone to sporadic oscillations in the solution if a grid with uniform spacing is used. This is achieved by casting Eq. (\ref{dade}) in Cartesian coordinates to obtain a numerical solution. Replacing the derivatives in Eq. (\ref{dade}) with their finite-difference counterparts yields
\begin{widetext}
\begin{eqnarray}
c_{i,j,k}^{n+1}=c_{i,j,k}^{n}+h_{\tau}\left[\left(\frac{1}{h^{2}}+\frac{\text{Pe}\;v_{x\;i,j,k}}{2h}\right)c_{i-1,j,k}^{n}+\left(\frac{1}{h^{2}}-\frac{\text{Pe}\;v_{x\;i,j,k}}{2h}\right)c_{i+1,j,k}^{n}+\left(\frac{1}{h^{2}}+\frac{\text{Pe}\;v_{y\;i,j,k}}{2h}\right)c_{i,j-1,k}^{n}+\right.\nonumber\\
\left.\left(\frac{1}{h^{2}}-\frac{\text{Pe}\;v_{y\;i,j,k}}{2h}\right)c_{i,j+1,k}^{n}+\left(\frac{1}{h^{2}}+\frac{\text{Pe}\;v_{z\;i,j,k}}{2h}\right)c_{i,j,k-1}^{n}+\left(\frac{1}{h^{2}}-\frac{\text{Pe}\;v_{z\;i,j,k}}{2h}\right)c_{i,j,k+1}^{n}-\frac{6}{h^{2}}c_{i,j,k}^{n}\right].
\end{eqnarray}
\end{widetext}

\par
Here $c_{i,j,k}^{n}$ is the value of the concentration at the location indexed by $i,j,k$ at the time indexed by $n$ with $h_{\tau}$ the time step and $h$ the distance between grid points. We work on a grid 10 units wide with 200 equidistant points along each axis. Here a unit corresponds to the dimensionless distance equal to one cell radius.  The average velocity field $\bm{\overline{v}}_{i,j,k}$ is computed first and stored in memory for each grid point.  The fluid model used to obtain our numerical solutions, the squirming swimmer, is defined in section IV. B. of \cite{felderhof2015stokesian}.  The only non-zero multipole moments are $\overline{m}_{1}=1/2$, $\overline{m}_{3}=-\overline{k}_{3}=7/48$, and $\overline{m}_{5}=-\overline{k}_{5}=-25/72$.  We use an explicit finite difference scheme where the concentration at time $n+1$ is computed from the concentration at time $n$.  This is continued until the percent change in the concentration field is no larger than 0.01\% per iteration at any grid point: indicating that the concentration field has reached steady state. The first-passage probability is then determined as the flux through the inner absorbing boundary divided by the sum of the flux through the inner and outer boundaries.
\begin{figure}[h]
	\centering
	\includegraphics[width=85mm]{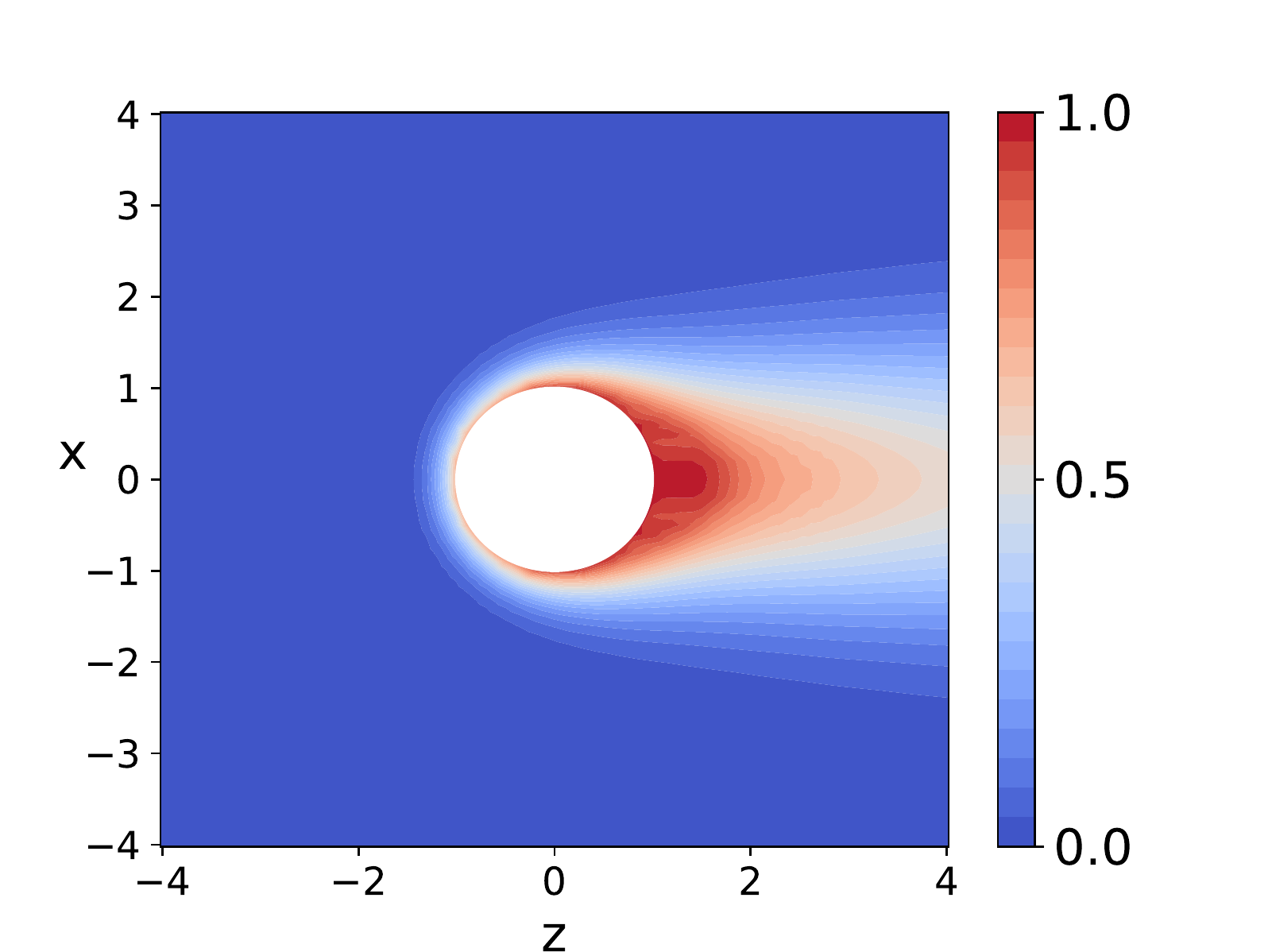}
	\caption{Numerical result for the first-passage probability with $\text{Pe} = 10$.  The characteristic plume structure shows that the far-field flow is directed from right to left in the figure, in accord with Eq. (\ref{flowfield}).   }
	\label{adfpp}
\end{figure}

\begin{figure}[h]
	\centering
	\includegraphics[width=85mm]{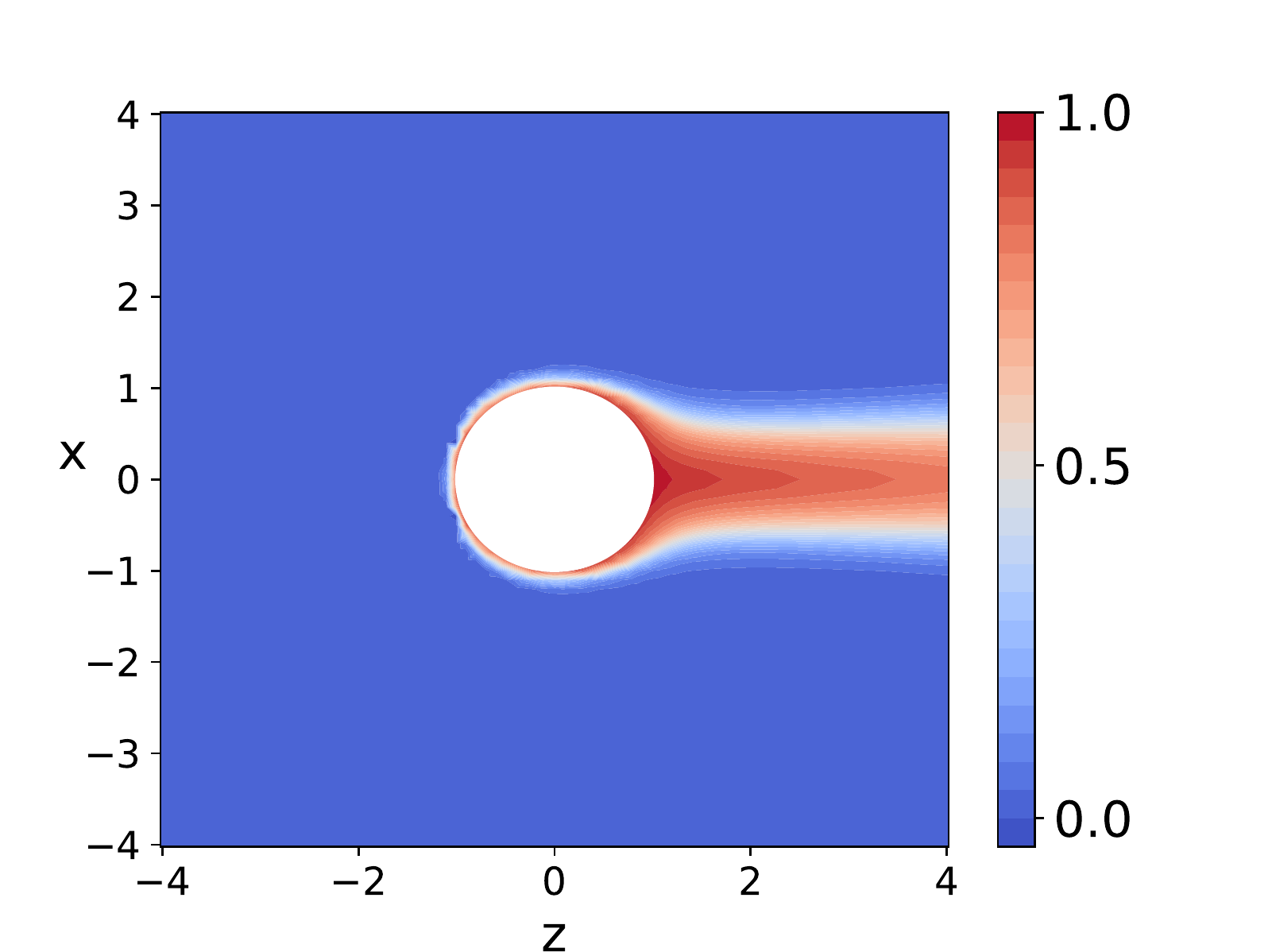}
	\caption{Numerical result for the first-passage probability with $\text{Pe} = 100$.  For larger Pe, notice the plume structure changes.  The thin layer of non-zero probability surrounding the cell provides a visual indication of the thickness of the concentration boundary layer.  }
	\label{adfpp}
\end{figure}
\section{Analytical Model}
\par
We now consider an approximate analytical approach to the problem.  The basic idea is to use methods common to boundary layer problems \cite{acrivos1960solution,acrivos1962heat,acrivos1965asymptotic,goddard1966asymptotic} to find the leading order solution of Eq. (\ref{dade}) with the boundary conditions given.  Readers familiar with this literature will recognize the general strategy, but our exposition is intended to be entirely self-contained so that a physics student with no previous exposure to the fluids literature but some exposure to boundary value problems (at the level common in an undergraduate electricity and magnetism or quantum mechanics course) can follow along.  The general structure for the calculation is as follows.  A first change of variables ($\xi \rightarrow \rho$) defines the boundary layer equation for our problem.  A second change of variables (the similarity transformation $\rho \rightarrow \eta$) gives us an effective radial equation.  This equation is then solved using methods similar to those students are familiar with from solving boundary value equations (Green's function).  
\par
Defining the small parameter $\alpha = 1/\text{Pe}$,
we first rescale the radial variable $\xi=1+\alpha^{n}\rho$ to stretch out the boundary layer.  The dimensionless time is rescaled as $\tau=\alpha^{m}T$, but we do not rescale the angular variables $\theta$ and $\phi$.  Dominant balance \cite{rnmg} determines the value of exponents $n=1/2$ and $m=1$.  By making this choice of exponents, in a perturbative expansion for the concentration, $c=\sum_{k=0}^{\infty}\alpha^{k/2}c_{k}$, the equation governing $c_{0}$ will contain temporal, advective, and diffusive terms. The physical thickness of the concentration boundary layer is $\ell=R\alpha^{1/2}$.
The fluid model is the same one used to obtain our numerical solutions.  Inserting the perturbative expansion into Eq. (\ref{dade}) and collecting terms of the same order in $\alpha^{1/2}$, the result is a system of coupled equations for the $\{c_{k}\}$.  Defining $\mu=\cos\theta$, the lowest order governing equation is
\begin{widetext}
\begin{equation}
\label{blex1}
\frac{\partial c_{0}}{\partial T}+\frac{105}{16}\rho (\mu - 6 \mu^{3}+5 \mu^{5})\frac{\partial c_{0}}{\partial \rho}- \frac{(1-\mu^2)}{32}(29-140\mu^{2}+175\mu^{4}) \frac{\partial c_{0}}{\partial \mu} -\frac{\partial^{2}c_{0}}{\partial\rho^{2}}=0.
\end{equation}
\end{widetext}

A common method in this type of boundary layer problem \cite{nuss,anu} is to define similarity variables $\eta=\rho/g$ and $\chi=T/g^{2}$, where $g(\mu)$ encapsulates the angular dependence of the boundary layer. Making this choice, the following relationships are needed for the change of variables: $\frac{\partial c_{0}}{\partial \rho} = \frac{1}{g}\frac{\partial c_{0}}{\partial \eta}$, $\frac{\partial^{2} c_{0}}{\partial \rho^{2}} = \frac{1}{g^{2}}\frac{\partial^{2} c_{0}}{\partial \eta^{2}}$, $\frac{\partial c_{0}}{\partial T} = \frac{1}{g^{2}}\frac{\partial c_{0}}{\partial \chi}$, $\frac{\partial c_{0}}{\partial \mu} = -\frac{\eta}{g}\frac{\mathrm{d}g}{\mathrm{d}\mu}\frac{\partial c_{0}}{\partial \eta}$.  This transformation isolates the angular dependence and defines the concentration boundary layer equation for the problem, 
\begin{widetext}
\begin{equation}
\frac{\partial c_{0}}{\partial \chi}+\eta \frac{\partial c_{0}}{\partial \eta} \left( \frac{105}{16}(\mu-6\mu^{3}+5\mu^{5})g^{2} + \frac{(1-\mu^2)}{32}(29-140\mu^2+175\mu^{4})g \frac{\mathrm{d}g}{\mathrm{d}\mu}\right)
-\frac{\partial^{2}c_{0}}{\partial\eta^{2}}=0.
\label{govzero}
\end{equation}
\end{widetext}
Provided the term in large parenthesis is equal to a constant $\Delta$, the equation is transformed into an effective radial equation.  To proceed we define the time-integrated concentration
\begin{equation}
\mathcal{C}_{0}=\int_{0}^{\infty}c_{0}\;d\chi.
\end{equation}
The equation for $\mathcal{C}_{0}$ becomes
\begin{equation}
c_{0}(\chi=\infty)-c_{0}(\chi=0)+\Delta \eta \frac{\partial\mathcal{C}_{0}}{\partial\eta}-\frac{\partial^{2}\mathcal{C}_{0}}{\partial\eta^{2}}=0. 
\label{radeq}
\end{equation}
Before solving this equation for the first-pasage probability, we discuss the solution to the associated angular equation for $g(\mu)$.  In what follows we make the choice $\Delta=2$, in which case we obtain a first-order differential equation for the variable $h=g^{2}$, 
\begin{equation}
\frac{\mathrm{d}h}{\mathrm{d}\mu} + p(\mu) h  =q(\mu)
\end{equation}
where
\begin{eqnarray}
p(\mu) &=& \frac{420 \mu (1-5\mu^{2})}{29 -140\mu^{2}+175 \mu^{4}} \\
q(\mu) &=& \frac{128}{29-169 \mu^{2}+315 \mu^{4} - 175 \mu^{6}}.
\end{eqnarray}
This equation can be solved with an integrating factor.  To do so we calculate $b(\mu) = \int p(\mu) \mathrm{d}\mu$ which gives
\begin{equation}
b(\mu) =  -6\sqrt{7} \tan^{-1}(\sqrt{7}(5\mu^{2}-2)) -3 \log(29-140\mu^{2}+175\mu^{4}).
\end{equation} 
The desired solution is then
\begin{equation}
h(\mu) = e^{-b(\mu)}\left(\int_{-1}^{\mu}q(s)e^{b(s)}\mathrm{d}s+K\right)
\end{equation}
where $K$ is an integration constant.  

We are now in a position to proceed with the calculatin of the first-passage probability.  Consider a spatial domain where all molecules released are eventually captured with probability one.  The concentration is defined in the spherical shell between two perfectly absorbing surfaces, the first at the surface of the cell $(\eta=0)$, and a second at some prescribed distance $(\eta=\eta_{+})$. 
Since all molecules are eventually absorbed, $c_{0}(\chi=\infty)=0$. If a molecule is released in the extracellular fluid at position $(\xi^{\prime},\theta^{\prime},\phi^{\prime})$, the corresponding initial condition is $c_{0}(\chi=0)=\delta^{3}(\bm{\xi} - \bm{\xi^\prime})$.  

Writing the initial condition in terms of $\eta$, our effective radial equation, Eq. (\ref{radeq}), becomes
\begin{equation}
\frac{\partial^{2}\mathcal{C}_{0}}{\partial\eta^{2}}-2 \eta \frac{\partial\mathcal{C}_{0}}{\partial\eta}= - \frac{\delta\left(\eta- \frac{g(\mu^{\prime})}{g(\mu)} \eta^{\prime}\right)\delta(\mu-\mu^{\prime})\delta(\phi-\phi^{\prime})}{\alpha^{1/2} g(\mu)(1+\alpha^{1/2} g(\mu)\eta)^{2}}.
\label{fpge}
\end{equation}
To solve Eq. (\ref{fpge}), note that the two independent solutions to the homogeneous equation
are a constant and $\text{erfi}(\eta)=\frac{2}{\sqrt{\pi}}\int_{0}^{\eta}e^{z^{2}}\mathrm{d}z $. Using perfectly absorbing boundary conditions at $\eta=0$ and $\eta = \eta_{+}$, the solution for $\mathcal{C}_{0}$ is
\begin{equation}
\mathcal{C}_{0}=\mathcal{I} \,\text{erfi}(\eta_{<}) \left( \text{erfi}(\eta_{>}) - \text{erfi}(\eta_{+}) \right).
\end{equation}
Here $\eta_{<}$ ($\eta_{>}$) is the smaller (larger) of $\eta$ and $\eta^{\prime}$. To determine the constant $\mathcal{I}$, we integrate both sides of the governing equation $\int_{0}^{2\pi}\mathrm{d}\phi\int_{-1}^{1}\mathrm{d}\mu \,g(\mu)  \int_{\eta^{\prime}-\epsilon}^{\eta^{\prime}+\epsilon} \mathrm{d}\eta$ to determine the discontinuity in the first derivative of $\mathcal{C}_{0}$,
\begin{equation}
\frac{\partial \mathcal{C}_{0}}{\partial\eta}\bigg|^{\eta=\eta^{\prime}+\epsilon}_{\eta=\eta^{\prime}-\epsilon}=-\frac{1}{2\pi G\alpha^{1/2}(1+\alpha^{1/2} g(\mu^{\prime})\eta^{\prime})^{2}}
\end{equation}
where $G = \int_{-1}^{1}\mathrm{d}\mu\, g(\mu)$.  The factor of $g(\mu)$ in the integration to determine the discontinuity was incorrectly omitted in \cite{urchin}, it is needed to ensure the proper normalization, 
\begin{equation}
\mathcal{I}=\frac{-e^{(\eta^{\prime})^{2}}}{4(\pi\alpha)^{1/2} G(1+\alpha^{1/2} g(\mu^{\prime})\eta^{\prime})^{2}}.
\end{equation}
\par
In terms of the original variables, the first-passage probability is calculated as
\begin{equation}
\Pi_{0}=\int_{0}^{\infty}\mathrm{d}\tau\int_{0}^{2\pi}\mathrm{d}\phi \int_{-1}^{1} \mathrm{d}\mu \,\xi^{2}\frac{\partial c_{0}}{\partial\xi}\bigg|_{\xi=1}.
\end{equation}
The result can be written in terms of the boundary layer variable $\eta$ and the time-integrated concentration $\mathcal{C}_{0}$ as
\begin{equation}
\Pi_{0}=2\pi\alpha^{1/2} \int_{-1}^{1} \mathrm{d}\mu\, g(\mu)\,\frac{\partial\mathcal{C}_{0}}{\partial\eta}\bigg|_{\eta=0}.
\end{equation}
Performing the angular integration we arrive at
\begin{equation}
\Pi_{0}=\frac{e^{-(\eta^{\prime})^{2}}}{(1+\alpha^{1/2} g(\mu^{\prime})\eta^{\prime})^{2}}\left(1-\frac{\text{erfi}(\eta^{\prime})}{\text{erfi}(\eta_{+})}\right).
\end{equation}
Moving the outer absorber $\eta_{+}$ out to infinity and writing the result back in terms of $\xi^{\prime}$ yields the final result,
\begin{equation}
\Pi_{0}=\frac{e^{-\frac{(\xi^{\prime}-1)^{2}}{\alpha h(\mu^{\prime})}}}{(\xi^{\prime})^{2}}
\label{zop}
\end{equation}
Note that the final result is properly normalized as $\Pi_{0}(\xi^{\prime}=1)=1$.  The result has certain structural similarities to a much simpler calculation, the first-passage probability in the case of pure diffusion, for which $\Pi_{\text{diffusion}}=1/(\xi^{\prime})^{2}$ \cite{redner2001guide}.  In words, the effect of fluid flow is an exponential supression of the first-passage probability as compared with the purely diffusive case.  
\section{Results}
\par
\begin{figure}[h]
	\centering
	\includegraphics[width=85mm]{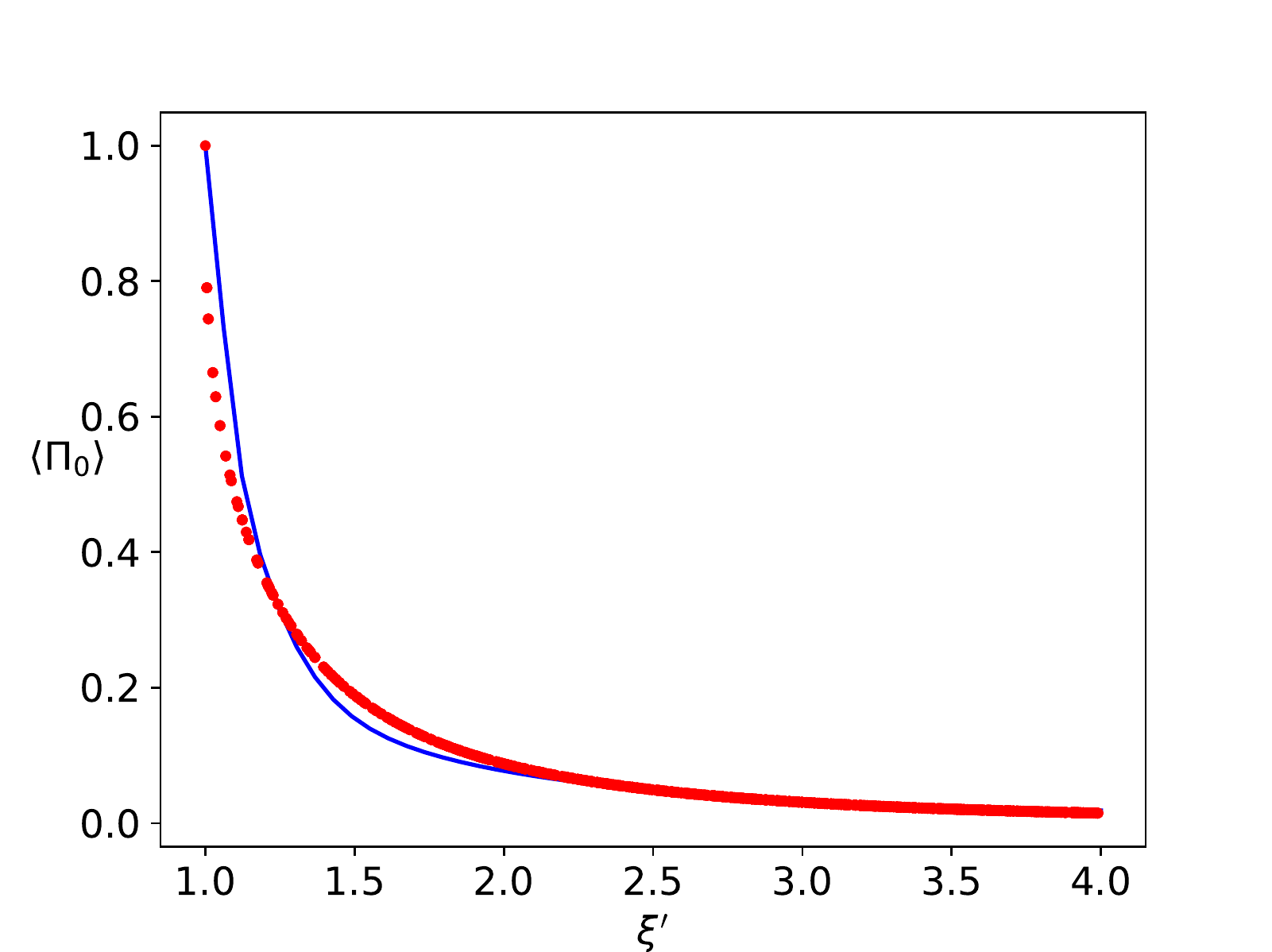}
	\caption{A comparison between the analytical results for the first-passage probability $\langle \Pi_{0} \rangle$ (blue line) and the numerical results $\Pi_{\text{fit}}$ (red dots) as a function of source position $\xi^{\prime}$ for $\text{Pe} = 100$.  The best fit parameters are $a=1.046$ and $b=0.2901$.}
	\label{adfpp100}
\end{figure}
\begin{figure}[h]
	\centering
	\includegraphics[width=85mm]{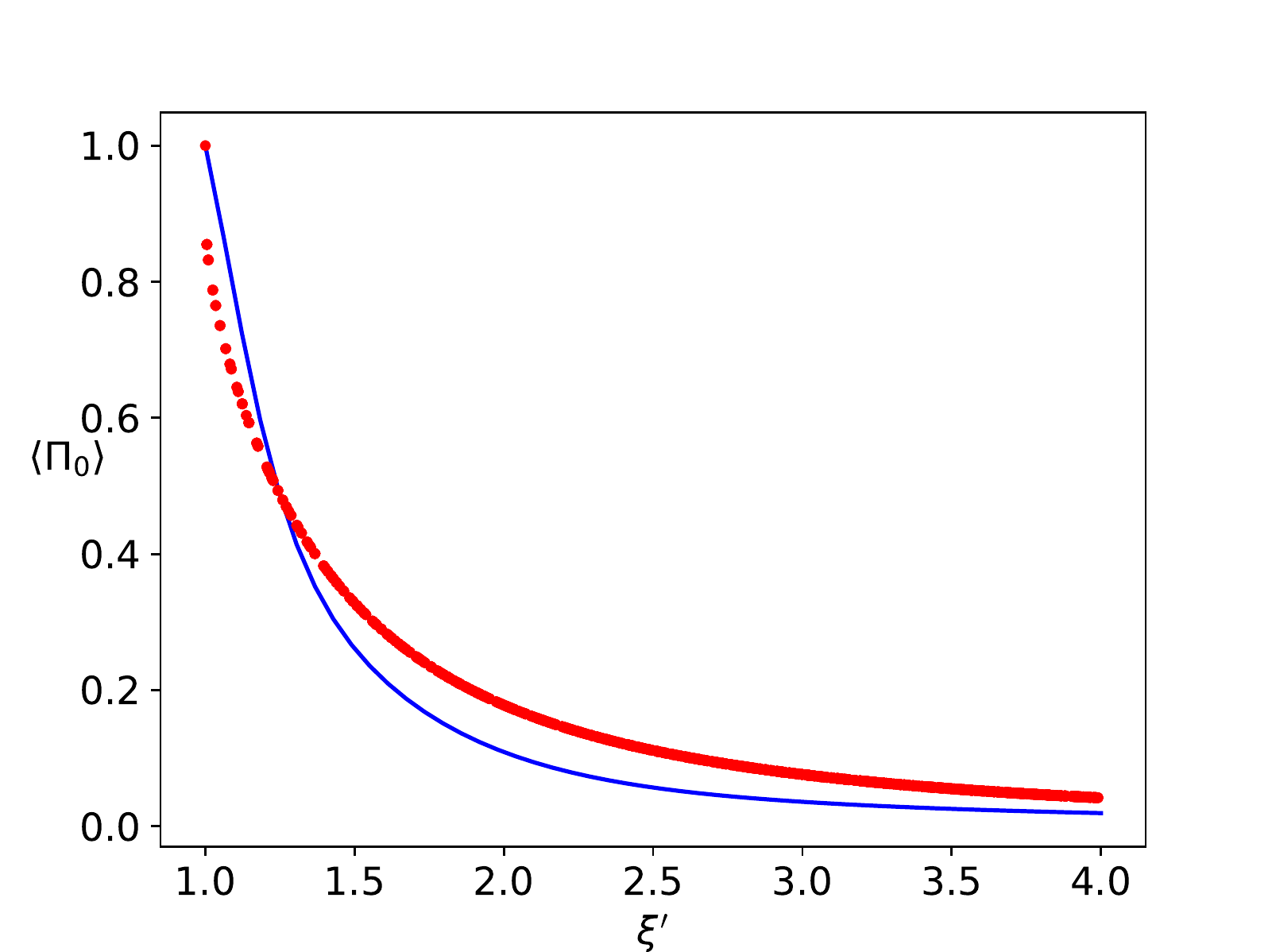}
	\caption{A comparison between the analytical results for the first-passage probability $\langle \Pi_{0} \rangle$ (blue line) and the numerical results $\Pi_{\text{fit}}$ (red dots) as a function of source position $\xi^{\prime}$ for $\text{Pe} = 10$.  The best fit parameters are $a=0.3394$ and $b=0.1582$.  }
	\label{adfpp10}
\end{figure}

To make a comparison between the analytical approach and the numerical solution for a variety of point source locations, we numerically calculate the average first-passage probability, $\langle \Pi_{0}\rangle = \frac{1}{2} \int_{-1}^{1} \mathrm{d}\mu^{\prime} \, \Pi_{0}$ from Eq. (\ref{zop}).  Here, the integration constant in the equation for $h(\mu)$ is zero, $K=0$.  The full set of angular results from the  numerical calculation are fit to a function of the form, 
\begin{eqnarray}
\Pi_{\text{fit}} = \frac{e^{-a(\xi^{\prime}-1)^{b}}}{(\xi^{\prime})^{2}}
\end{eqnarray}
using a non-linear least squares routine in Python.   
Figures \ref{adfpp100} and \ref{adfpp10} show a comparison of the numerical result for the first-passage probability and the analytic approximation.  The result for the first-passage probability is in good agreement with the numerical result, especially at large Pe (see Fig. \ref{adfpp100}).  This is reassuring as the perturbation program is constructed based on a small parameter Pe$^{-1/2}$.  As Pe is reduced the agreement becomes less quantitatively accurate (see Fig. {\ref{adfpp10}}), but contintues to capture the qualitative behavior, even for source locations far outside of the concentration boundary layer.  \\
\par To make contact with the motivations for the study discussed in the introduction, note that the cost and effectiveness of the toxin transport system is an active area of experimental research \cite{cole2013cost}.  A physical microvilli length of $4 \,\mu$m corresponds to a source location of $\xi^{\prime} = 1.1$ in Figs.  \ref{adfpp100} and \ref{adfpp10}.  For a cell in a flow field (either generated by its own swimming motion or an environmental flow), designing a {\it smart} toxin disposal system would probably only require that the cell have knowledge of the upstream direction.  In the biophysical context, the receptors responsible for effluxing toxin molecules to the exerior of the cell do not chemically modify the toxin.  As a result, there is a potentailly costly scenario in which effluxed molecules are reabsorbed and have to be discarded again, which is known as futile cycling.  The cell could see significant gains in efficiency be either preferentially activating transport receptors downstream or actively transporting molecules tagged for export to downstream receptors.  A future experiment designed to monitor receptor activity in a controlled flow environment, perhaps a microfluidic chamber, might be able to uncover whether a smart disposal mechanism similar to this has evolved naturally.  
\par
\section{Conclusion}
Motivated by a mass transport problem during embryonic development, we considered an active particle model for a swimming cell, the spherical squirmer.  Within the context of the model, we determined the first-passage probability for an advection-diffusion equation at small Reynolds number and large P\'eclet number.  Numerical approaches to solving the advection-diffusion equation based on explicit finite-differencing were compared to an analytic approximation based on a rescaling of the concentration boundary layer.  For large P\'eclet number we find quantitative agreement between the approaches.  As the P\'eclet number is reduced, the analytic approximation continues to capture the qualitative behavior of the first-passage probability, but deviates somewhat from the numerical results.  The regime of validity for the analytic approximation might be improved by continuing the perturbation program beyond the zeroth order.  This is a potential direction for future research.  

\section{Acknowledgements}
This work was supported by faculty startup grant U035843 from the College of Arts, Sciences, and Letters at the University of Michigan-Dearborn.

\bibliographystyle{ieeetr}
\bibliography{ref}

\end{document}